\documentclass[useAMS,usenatbib]{mn2e}
\usepackage{url}
\usepackage{graphicx}
\newcommand{\doo}[2]{{\frac{\partial #1}{\partial #2}}}
\newcommand{\fat}[1]{{\bf #1}}

\newcommand{\eps}{{\epsilon}}
\newcommand{\cross}{ {\fat \times } }
\newcommand{\commentti}[1]{ {~} }
\newcommand{\emato}{{ \widetilde e}}
\begin{document}

\title{Quasi-satellite dynamics in formation flight}

\author[Mikkola and Prioroc]{Seppo Mikkola$^1$\thanks{E-mail:seppo.mikkola@utu.fi}~\&~
Claudiu-Lucian Prioroc$^{1,2}$\thanks{E-mail:cprioroc.ext@gmv.com}\\
              $^1$Department of Physics and Astronomy, University of Turku,\\
                Tuorla Observatory, V\"ais\"al\"antie 20, Piikki\"o , FI 21500, Finland \\
$^2$GMV, Calea Victoriei 145, Bucuresti 010072, Romania}

\date{\today}

\pagerange{\pageref{firstpage}--\pageref{lastpage}} \pubyear{}

\maketitle

\label{firstpage}

\begin{abstract}
The quasi-satellite (QS) phenomenon  makes two celestial bodies to fly near each other \citep{StabLim}
and that effect can be used also to make artificial satellites move in tandem.
We consider formation flight of two or three satellites in low eccentricity near Earth orbits.
 With the help of weak ion thrusters it is possible to accomplish tandem flight.
With ion thrusters it is also possible to mimic
many kinds of mutual force laws between the satellites. We found that
both a constant repulsive force or an attractive force that decreases with
the distance are able to preserve the formation in which the eccentricities
cause the actual relative motion and the weak thrusters  keep
the mean longitude difference small.
Initial values are important for the formation flight but very exact adjustment
of orbital elements is not important.
Simplicity is one of our goals in this study and this result is achieved
at least in the way that, when constant force thrusters are used,
the satellites only need to detect
the directions of the other ones to fly in tandem.
A repulsive acceleration of the order of $10^{-6}$ times the Earth attraction, is enough
to effectively eliminate the disruptive effects of all the perturbations at least for a timescale of years.
\end{abstract}


\section{Introduction}
 In an attempt to understand better the quasi-satellite phenomenon \citep{StabLim} we tested what would happen
if the force between two co-orbital bodies were different from the normal Newtonian gravity.
We found that many forces can produce similar relative motion of the bodies.
Even repulsive forces result in tandem motion of the bodies. At that point it became clear that
constant weak repulsive force could be used to keep artificial satellites flying near each other.
This paper studies that phenomenon mainly numerically but we also present some simple analytical
considerations.

In quasi satellite motion
an asteroid moves around the Sun co-orbitally with a planet and
 remains near the planet such that in the rotating coordinate system
it looks like moving around the planet in a retrograde orbit.
In that system the mutual force of the bodies is attractive. On the other hand
it is well known that a spacecraft chasing a satellite in the same orbit must brake to catch it up.
Thus, somewhat counter intuitively,
tandem flight of two satellites moving around a central body in its gravitational field
seems to be possible both with attractive or repulsive mutual acceleration.
In the case of two satellites the force between the satellites can been mimicked
using ion thrusters. Both an attractive force $\propto 1/\Delta^2$
 and repulsive constant force produce similar effects, which may look somewhat unexpected.
Above $\Delta$ means the distance of the satellites which we assume to be quite small
compared with the size of the orbit. We restrict our consideration to low eccentricity orbits.
High eccentricity orbits are generally more complicated for formation flying \citep{HighEcc}.

The simplest of all methods to make satellites to fly
in tandem is to put them into precisely same orbit to some distance from each other.
{An example of such a real system is NASA's Grail mission in which two spacecrafts are flying in tandem orbits
around the Moon to measure its gravity field in detail.}

This works perfectly as long as one can consider the gravitational field to be rotationally
symmetrical (like the Earth's field modeled without tesseral harmonics) in which case
the distance between the satellites varies only due to eccentricity caused speed differences.
The simplest way to achieve tandem flight in that situation
is possibly making ion thrusters to point towards, or away from, the other satellite(s) and this seems to work
according to our simulations. One can keep the satellites to move in the same fashion as the natural
quasi-satellites. Even near triangle configuration for three satellites is possible without
any complicated control, just constant acceleration or an acceleration that depends on the distance of the satellites.
No other information is necessary but the relative directions (and distances  if the acceleration needs to depend
on the distance).
{Long lasting thrusters are necessary, but such exist as shown
e.g. by ESA's SMART-1 mission from a near Earth orbit to the Moon.} 

The very basic method to consider satellite relative motions near each other and/or rendezvous is naturally the
Clohessy-Wiltshire equations \citep{CloWil}.
To get satellites flying in tandem has been considered in numerous publications e.g.
precise orbital elements adjustments have been discussed in  e.g. \citep{AlfEtAl,BGMR2004,ShaubAlf2,DiffEquinoctial}.

\cite{KPLR} considered the formation flight using differential approximations of the pure Kepler motion, { also \cite{C-LPSM}
discuss the computation of relative motion using simple methods. \cite{HighEcc} discuss formation flight in highly eccentric
orbits using differential mean orbital elements as design variables and
the \cite{DiffEquinoctial} state transition matrix for relative motion propagation. Use of a coordinate transformation
was consider by \cite{VA14}.}

We first consider the problem in terms of a simple approximate analytical theory. The assumptions in that  theory
are based on simulation results since a complete analytical solution seems laborious.
After the analytical considerations we present results from numerical simulation.
The gravity field that we used in the tests was the spherical function expansion, following \cite{MPH02},
plus Luni-Solar terms, but we also studied  the simple $J_2$ field model.
The ion thruster effects we model using different model potentials
could be engineered with adjustable thrusts. However, the constant force model is the simplest one
and it seems to work.

Finally we remark that this is not an engineering paper, we just study the dynamics.

\section{Analytical considerations}

\subsection{Units and orbital elements}
We choose the units such that
$$
\begin{array}{cl}
 G=1,~
R_\oplus =1,~
m_\oplus=1
             \end{array}
$$
where $G$=the gravitational constant, $R_\oplus$=radius of the Earth
(actually 6378.137~km) and $m_\oplus$= the mass of the Earth. The unit of time becomes such that the formal
period of a satellite with semi-major-axis $a=1$ would be $=2\pi$. In this system  one time unit in seconds
is $=806.811064922699$~sec and the length of a day is $=107.088$.
Despite these choices we write in some formulae the radius of the Earth and its mass explicitly visible.
In illustration, to make them easier to understand, we  use also other units, e.g. kilometers.
\newcommand{\inc}{{i}}
For orbital elements we use the standard notations
$$
a,\;e,\;\inc,\;\omega\;,\Omega,\;M,
$$
which are the semi-major axis, eccentricity, inclination, argument of perihelion,
longitude of the ascending node and the mean anomaly, respectively.
For the position vector we use $\fat r$ and for velocity $\fat v$. In the used units the velocity and momentum are the same i.e. $\fat v=\fat p$.
The semi-major-axes of the satellites are only very slightly
different (and equal on the average)  and therefore we can in most consideration take them to be practically same although the
difference $\delta a$ has an important role in some analytical considerations.

We need also the difference of the mean longitudes $\lambda=M+\omega+\Omega$.
We consider (almost) coplanar orbits which means that $\Omega_1=\Omega_2$
to high precision and therefore the difference of $\lambda$'s is essentially the same as
the difference of the angles $M+\omega$. This quantity is one of the most important ones in
our analytical consideration of the stability of the tandem flight. We use for it the notation
\begin{equation}
\theta=\lambda_2-\lambda_1=M_2-M_1+\omega_2-\omega_1,
\end{equation}
where $\dot M_k=n_k=1/\sqrt{(a_k^3/m_\oplus)}$ are the mean motions.
 An other important quantity is the
eccentricity vector
\begin{equation}
\fat e=\fat v\cross (\fat r\cross \fat v)/m_\oplus-\fat r/r,
\end{equation}
and especially their difference
\begin{equation}
\widetilde{\fat e}=\fat e_2-\fat e_1,
\end{equation}
which is a kind of relative eccentricity vector.

\subsection{C-W/Hill theory}
\cite{CloWil} presented a theory of satellite rendezvous (C-W hereafter).
This theory used the variational equations of the two-body problem
\begin{equation}\label{variaatio}\label{cw-eq}
\dot {\fat x}=\fat w,\ \ \ \dot {\fat w}=-m_\oplus (\fat x/r^3-3\fat r\cdot \fat x\: \fat r/r^5),
\end{equation}
where $\fat x$ is the variation of the position and $\fat w=\dot {\fat x}$ is the variation of velocity.
For a circular orbit in the rotating coordinate system, where $\fat r$ is constant,
 the equations become linear and easily solvable.
Much earlier Hill had studied the Lunar orbit in a somewhat similar way \citep{Hill1878}.
The difference in these treatments is that in the latter one there are additional 
nonlinear terms in the equations (\ref{cw-eq}) of motion. 
As shown e.g. by  \cite{C-LPSM} the C-W-solution and in fact a somewhat more general one 
can  be obtained by differentiating the two-body equations with respect to the orbital elements.  
Let $q_k\ \ k=,1,2..,6$ be a set of elements e.g.
$
\fat q=(M_0,i,\Omega,\omega,a,e),
$
which are the  mean anomaly $M$ at epoch, inclination, ascending node, argument of pericentre, semi-major-axis and eccentricity, respectively.
The solution for the variational equation
takes the form
\begin{equation}
 \fat x=\sum \doo{\fat r}{q_k}\delta q_k \ \ {\rm and} \  \fat w= \sum\doo{\fat v}{q_k}\delta q_k.
\end{equation}
The constants $\delta q_k$ represent the element variations. A particular solution of equation~(\ref{variaatio})
is 
\begin{equation}
\fat x_a=2\fat r-3\: t \: \fat v,
\end{equation}
which is related to the variation of the semi-major-axis. Because the time $t$ appears as a factor of $\fat v$
it is clear that
without some regulating additional force the distance between the bodies would
increase continuously  unless the semi-major-axis difference is precisely zero.

Additional information about the stability of the Hill-type  and
distant retrograde  orbits  can also be found in
\cite{Jackson}, \cite {sheeres} and \cite{StabLim}.

\subsection{Secular dynamics}

\subsubsection{ Assumptions based on numerical results}
Here we present some numerically obtained facts that can be used as
assumptions in our analytical considerations. The force model we used
consisted Earth potential model with the $J_2$ terms and a weak
ion thruster acceleration that was either a repulsive constant acceleration
or, for comparison, an attraction that had the $1/\Delta^2$ distance dependence.
Also other models were briefly consider (see below).

The following was found to be true in our simulations:
\begin{description}
\item
-The orbital elements vary periodically but their mean values remain the same
and the $J_2$ (and higher order) effects have minor influence in the relative
motion of the satellites (see also \cite{C-LPSM}).

\item
-Inclinations $i$ and longitudes of ascending node $\Omega$ remain very
close to each other if they are so initially. This is due to
similar precession rates.
Thus the satellites remain almost co-planar.  Also the $\omega$'s precess at the same rate.
\item
-The difference of the eccentricity vectors $\widetilde{\fat e}$ remains almost constant
in length. This is important in the theory and in accordance with the results
obtained in \cite{StabLim}. See Fig.\ref{eeekuva} in this paper.
\item
-When the initial semi-major-axes are nearly the same they remain
so as long as the satellites fly in tandem. Due to the thruster effect
the values fluctuate such that the mean angular speeds remain the same.
Thus in an approximate theory
we can take the (mean) semi-major-axes to be the same, only their difference varies periodically.
\end{description}

Those observations can be used as assumptions in the next
section where the analytical treatment is discussed.

\subsubsection{Analytical approximations}
Consider first a simple case in which we assume the motions to be pure
two-body motions and of low eccentricity. Further we assume the motions to be co-planar.

As a first simple example we consider the case in which one of the orbits
is circular and the other one has a small eccentricity. In the rotating
coordinate system in which the circular orbit satellites coordinates
are simply $=a(0,1)$ when the distance is consider to be the $y$-coordinate,
the eccentric orbit has approximately the coordinates
\footnote{The corresponding formulae in \cite{StabLim} are in error but not used.}
$$a(2 e \sin(M)+\theta,1-e \cos(M)),$$
when eccentricity and $\theta$ are
consider to be quantities of the same order of magnitude
 {(we call this order $O(\varepsilon)$)} and so small that
the linearized approximation is valid.

\begin{figure}
\begin{center}
{\includegraphics[width=8cm,height=4.7cm]{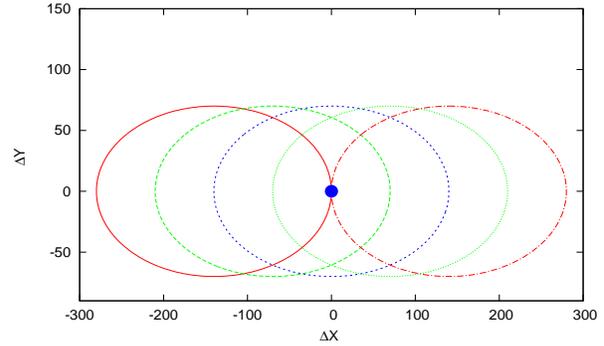}}
\caption{The relative motion of the satellites for the cases when
$\theta=\pm 2 e$, $\theta=\pm  e$ and $\theta=0$ in the coordinate
system that rotates with one of the satellites(=the central dot=satellite~1).
Here the  eccentricity of satellite 2 is $e=0.01$,
while the satellite 1 has eccentricity 0.
The distances were converted to kilometers assuming that $a=7000\:$km.
\label{kuva1}}
\end{center}
\end{figure}
 The differences of the coordinates of the satellites
are thus
\begin{equation}
(\Delta x,\Delta y)=a(2 e \sin(M)+\theta,-e \cos(M))+\vec O(\varepsilon^2),
\end{equation}
from which we see that the relative motion is mainly caused by the eccentricity as long as
the mean longitude difference $\theta$ remains small. The effect of $\theta$ is illustrated in Fig.\ref{kuva1}.
From the above coordinate differences we have for the distance
between the satellites the form
\begin{equation}\label{simpled}
\Delta^2= a^2((\theta+2 e \sin(M))^2+e^2 \cos^2(M))+O(\varepsilon^2).
  \end{equation}
The generalization of this to the case of two eccentric orbits is quite
simple (see below).

For stable tandem flights it is clearly necessary that the mean longitude difference $\theta$
indeed remains small. This is what we  study now.

If both satellites are supposed to use the ion thrusters
with equal accelerations, it is possible to
model the system mathematically in terms of a Hamiltonian

\begin{equation}
H=K_1+K_2-R_1-R_2-R_{12},
\end{equation}
where
\begin{equation}
K_\nu=\fat p_\nu^2/2-\frac{m_\oplus}{r_\nu},
\end{equation}
and  $R_k$'s are the perturbing force functions  with high
order spherical harmonics, including tesseral ones and the rotation of the Earth, were included. The force function for both satellites was
\small
\begin{equation}\label{Usarja}
R=\sum_{n=1}^{N_x}\sum_{m=0}^n \frac{P_{n}^m(\cos(\theta))}{r^{n+1}}(C_n^m\cos(m\psi)+S_n^m\sin(m\psi))+R_{ls}.
\end{equation}
\normalsize
Here $\psi=\phi-n_E t-\eta_0$, $\phi$ is the longitude, read eastwards, in the Earth fixed coordinates,
$t$ is time, $n_E$ is Earths rotational frequency,
$\eta_0$ initial phase, while $N_x$
defines the order up to which terms are taken into account (we used $N_x=36$, like \cite{MPH02}).
The numerical coefficients $C_n$ and $S_n$ define the gravitational potential.
$P_{n}^m$'s are the associated Legendre polynomials. The term $R_{ls}$ naturally signifies
the effect due to the Sun and Moon (ls = Luni-Solar).

If we include only the leading term, the $J_2$ term, then
\begin{equation}
R_\nu=\frac{m_\oplus}{r_\nu}(\frac{J_2R_\oplus^2}{2r_\nu^2}(1-3(\frac{z_\nu}{r_\nu})^2),
\end{equation}
are the $J_2$ terms ($J_2=0.0010826299890519$ in this notation).
 The `perturbing' thruster acceleration is in the term $R_{12} = R_{12}(\Delta)$
(which thus depends only on $\Delta$), where
\begin{equation}
 \Delta=|\fat r_2-\fat r_1|.
\end{equation}
The secular  (averaged over one period) form of the $K_\nu$ Hamiltonians in the $J_2$ approximation are
\begin{equation}
 \langle K_\nu-R_\nu\rangle=-\frac{m_\oplus}{2 a_\nu}-\frac{{J_2R_\oplus^2} m_\oplus \left(2-3 \sin ^2(i_\nu)\right)}{4 a_\nu^3 \left(1-e_\nu^2\right)^{3/2}.}
\end{equation}
The perturbing function $R_{12}$ tells how the thrusters are used.
For example if constant acceleration is used, then
\begin{equation}
R_{12}(\Delta)=\eps\Delta,
\end{equation}
where the constant $\eps$ is the satellites acceleration due to the ion thruster.
We assume this to be constant (unless otherwise indicated). 

Following \cite{StabLim} for co-planar orbits one can derive
the approximation
\begin{equation}\label{deltamato}
\Delta^2\approx a^2 (\emato^2\cos(w)^2+(\theta+2\emato \sin(w))^2)
\end{equation}
where
\begin{equation}
\emato=|\fat e_2-\fat e_1|,\  \   \ w=M+{\rm constant}.
\end{equation}
Thus this more general approximation differs from
the simple equation (\ref{simpled}) only in the way that the eccentricity
is replaced by the `relative eccentricity' $\emato$ and the
angle $w$ differs from $M$ by a constant (of two-body motion).

\subsection{Equations of motion}
The secular perturbing function for the two-satellite motion is
\begin{equation}
R=\sum_{\nu=1}^2\frac{{J_2} m_\oplus R_\oplus^2 \left(2-3 \sin ^2(i_\nu)\right)}{4 a_\nu^3 \left(1-e_\nu^2\right)^{3/2}}+\eps <\Delta>.
\end{equation}

The usual Lagrange equations for the orbital elements
 $a,\;e,\; i,\; \varpi,\;\Omega$ and mean anomaly $ \; M$
can be written in the form 
\begin{eqnarray}
\dot a&=&\frac{2}{na}\doo RM \;\;\; ,\;\;\;\;\;\nonumber\\
\dot e&=&\frac{-\sqrt{1-e^2}}{na^2e}\doo R{\omega}+
        \frac{1-e^2}{na^2e}\doo RM  \nonumber\\
\dot i&=&\frac{-1}{na^2\sqrt{1-e^2}\sin (i)}\doo R{\Omega}+
        \frac{\cos(i)}{na^2\sqrt{1-e^2}\sin (i)}\doo R{\omega} \nonumber   \\
\dot {\Omega}&=&\frac{1}{na^2\sqrt{1-e^2}\sin (i)}\doo Ri   \nonumber\\
\dot {\omega}&=&\frac{\sqrt{1-e^2}}{na^2e}\doo Re
-\frac{\cos(i)}{na^2\sqrt{1-e^2}\sin (i)}\doo Ri   \nonumber\\
\dot M&=&n-\frac{2}{na}\doo Ra -\frac{1-e^2}{na^2e} \doo Re   \\
\end{eqnarray}
Since the semi-major-axes $a_1$ and $a_2$ are equal on the average, we
write $a$ for this average, and $n=1/(a\sqrt{a/m_\oplus})$ for the corresponding mean motion.

The derivative of $\theta$ reads
\begin{equation}
\dot\theta=\dot\lambda_2-\dot\lambda_1=\dot M_2+\dot\omega_2-\dot M_1-\dot\omega_1.
\end{equation}
Due to the smallness ($O(J2)$) and similarity ($ \dot\omega_2\approx\dot\omega_1$)  of the perturbations the main term in the equation for $\dot \theta $
can be taken as
\begin{equation}
\dot \theta\approx n_2-n_1\approx -\frac32 n \alpha
\end{equation}
where $\alpha=(a_2-a_1)/a$ and because of the expression (above) for $\dot a$ and the fact that $M$ is present only
in $\theta=M_2-M_1+\omega_2-\omega_1$, we have
\begin{equation}\label{tupladot}
\ddot \theta\approx-\frac{3}{a^2}(\doo{R}{M_2}-\doo{R}{M_1}) = -\frac{6}{a^2}\doo{R}{\theta},
\end{equation}
in which the last equality follows from the fact that $\doo{R}{M_1}=-\doo{R}{M_2}=\doo{R}{\theta}$.
Note that the inclusion of the $\doo{R}{M_1}$-term is due to our assumption that both satellites have
a thruster. At this point one sees that the $J_2$ term has no effect in the motion of $\theta$.
However it has a small indirect effect because the orbits are not Keplerian and the secular
theory is not accurate enough. These facts are illustrated in Fig.~\ref{kuva2} in which
one can see the effect of $J_2$ as compared with the pure two-body motion. In these experiments the thrusters
perturbed the satellite motions in the same way.

In addition we really have the term $\ddot \omega_2-\ddot\omega_1$, but here the terms nearly cancel each other
and they are very small =$O(J_2^2)$, so that the equation (\ref{tupladot}) is a good approximation.

The secular perturbation due to the thruster effect
can now be obtained by averaging $\Delta$ over the angle $w$ as
 \begin{equation}
\langle R_{12}\rangle =\frac{1}{2\pi}\int_0^{2\pi}R_{12}(\Delta(w))\: dw
 \end{equation}
As said above, for constant mutual force $R_{12}=\eps \Delta$.
Due to the structure of the expression for $\Delta^2$ it is clear that
$\langle R\rangle $ is an even function of $\theta$, i.e.
\begin{equation}\label{Rexpansion}
\langle R_{12}\rangle =R_{12}^{(0)}+R_{12}^{(2)} \theta^2/2+..,
\end{equation}
Averaging over the angle $w$ in Eq.(\ref{deltamato}) and expanding in powers of $\theta$ one gets the secular approximation
\begin{eqnarray}
{\langle R_{12}\rangle } &\approx& \eps a (1.542 \emato + 0.1426\theta^2/\emato+O(\theta^4)),
\end{eqnarray}
These equations are naturally valid only to second order in $\theta$ since that is the case
for Eq.(\ref{deltamato}).
For the equation of $\theta$-motion we have
\begin{equation}\label{thetaeqs}
\ddot \theta=-\frac{6}{a^2} \doo{R_{12}}{\theta}\approx \frac{-1.711 \eps}{a \emato}\theta,
\end{equation}
which shows that the $\theta$-motion is harmonic oscillation in this approximation.
In Fig.~\ref{theta.periods} the periods are compared for the  different  models ( using Eq.(\ref{Usarja}),  the $J_2$ only  model
and pure two-body motion) and the results are quite close to each other, with only a few percent differences. Thus one may conclude that
the $R_{12}=\eps \Delta$ perturbation mainly determines the motion of the mean longitude difference $\theta$.

In general it is interesting to consider the effect of different
kinds of mutual interactions. For a general power $n$ of the distance
one may write the average of $R=\eps_n \Delta^n$
as
\begin{equation}
\langle R\rangle=\eps_n \frac{1}{2\pi}\int_0^{2\pi}\Delta^n(w)\: dw.
\end{equation}

\begin{figure}
\centerline{\includegraphics[width=8cm,height=4.5cm]{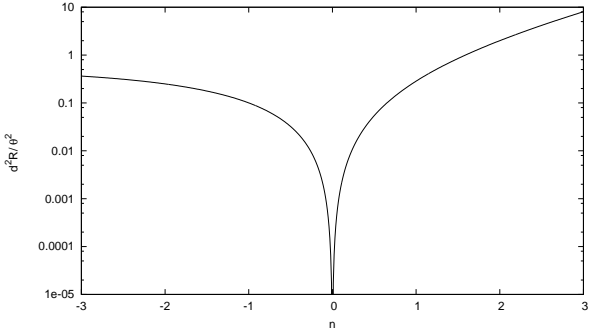}}
\caption{ Second derivative of $e^{2-n}\langle \Delta^n\rangle $ with respect to $\theta$ (at $\theta=0$) as function of the power $n$.
One sees that it is positive for all non-zero powers $n$, thus
predicting possibility for stable motion near each other.\label{ddthn}\label{kuva7}
}
\end{figure}

The sign of the second derivative of $\langle R\rangle $, with respect to $\theta$, tells if the
mean longitude difference $\theta$ can stay small.  In fact is turned out, somewhat unexpectedly,
that the second derivative is a positive number for all nonzero powers $n$, positive or negative.
This result is illustrated in Fig.\ref{ddthn}.
Thus the tandem flight seems possible with any mutual acceleration of the satellites
if the force is derivable from a potential of the form $\eps_n \Delta^n$ for ($n\ne 0$) where $\eps_n>0$.
In passing we mention that the perturbing function can in general be anything of the form
$R=\sum_{n=-\infty}^{+\infty} c_n \Delta^n$,
{which may be a finite, or at least convergent, expression still giving positive second derivative [$R_{12}^{(2)}$ in (\ref{Rexpansion})] at $\theta=0$.
This happens at least if the coefficients $c_n$ are non negative.}

\section{Numerical experiments}

In this section we first discuss the results with two-satellite experiments and later this is extended to three
satellites moving essentially in a (flattened) triangle configuration.

Our numerical experiments confirm clearly the main point of the theory part: the mean longitude difference $\theta$
fluctuates around zero and remains small. The amplitude depends on the difference in the initial values of the
semi-major-axis, but the effect of the thruster makes this difference fluctuate around zero according to the
equation (\ref{thetaeqs}) and so the mean values are same. Due to the fact that the two orbits are very
similar, all the perturbations in them are similar and do not affect noticeably the relative motion of the satellites.

The initial values in the numerical experiments were produced using two-body formulae. We set initially
the inclinations $i= 10^\circ, 30^\circ, 50^\circ, 70^\circ, 90^\circ$  and longitudes of ascending nodes $\Omega$ to same value for the satellites.
For the semi-major-axes we used typically
one kilometer  difference in the initial axis value. The initial
eccentricities were  set to the values of $e= 0.01$.
The differences in the initial positions of the satellites were obtained by adding $180^\circ$ to the value of $\omega$ and subtracting
that amount from the mean anomaly $M$.
For the experiments with three satellites we used
 $(k-1)*120^\circ$ ($k=1,\ 2,\ 3$)
$\omega$ differences and removed an equal amount from $M$. These operations made the mean longitudes equal in the beginning.

\subsection{Two satellites}
\begin{figure}
\centerline{\includegraphics[width=8cm,height=4.5cm]{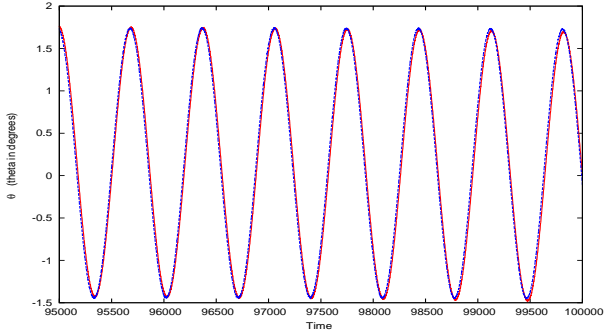}}
\caption{Oscillation of the angle $\theta$ in two models: with the entire
Earth potential and with $J_2$-term only. After about 2.5 years (which corresponds near the middle of this
figure, in which the total time-span is about 1.5 months) the oscillations are still almost the same within plotting precision. Here $i=30^o$, $\delta a_0=1km$.
\label{theta_in_two_models}\label{kuva2}}
\end{figure}
\begin{figure}
\centerline{\includegraphics[width=8cm,height=4.5cm]{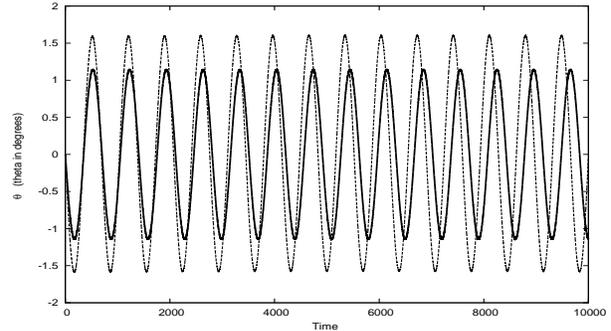}}
\caption{Oscillation of the angle $\theta$ in two models: with the entire
Earth potential and with pure two-body model. Here the time-span is about three months
and the inclination and semi-major-axis difference are correspondingly
 $i=30^o$, $\delta a_0=1km$.\label{kuva3}
}
\end{figure}
\begin{figure}
\centerline{\includegraphics[width=8cm,height=4.5cm]{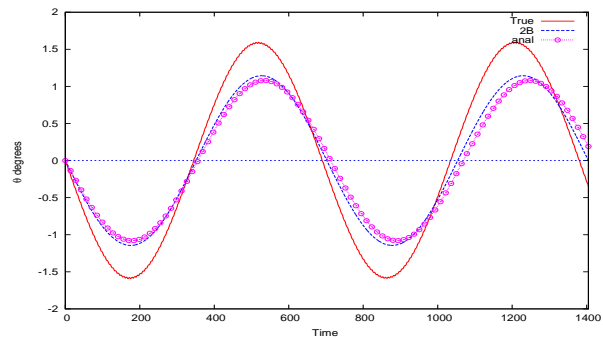}}
\caption{In the different approximations the $\theta$ periods are somewhat different,
but no more than one could expect. The 'True' curve is from our most accurate
modelling (all perturbations included), the '2B' means two-body motion model
without any of the Earth harmonics and 'anal' points to the result computed using the
above equation  (\ref{thetaeqs}). The differences in the period length are only a few percent.
\label{theta.periods}\label{kuva4}}
\end{figure}

\begin{figure}
\centerline{\includegraphics[width=8cm,height=4.5cm]{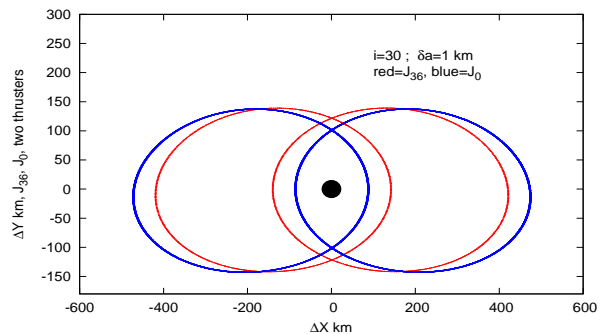}}
\caption{Comparison of relative orbits at extreme $\theta$-values for the case
of full perturbations (red) and in the two-body model.
 Here $i=30^o$, $\delta a_0=1km$.
\label{kuva5}}
\end{figure}

\begin{figure}
\centerline{\includegraphics[width=8cm,height=4.5cm]{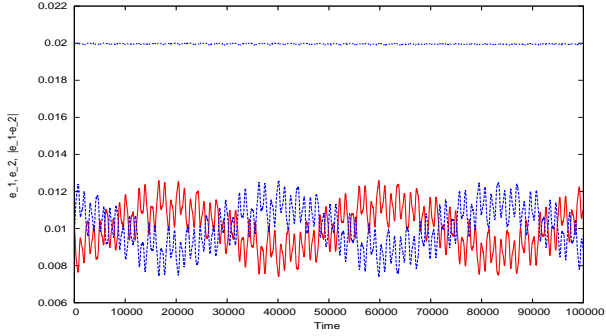}}
\caption{Evolution of individual eccentricities and the relative eccentricity.
While the individual eccentricities $e_k$ fluctuate considerably, the relative
one  $\widetilde e=|\fat e_1-\fat e_2|$ remains almost constant. In this example
the initial semi-major axis difference was one kilometer, $e_k=.01$ and $\omega$-difference
was set to be $=180^o$.\label{eeekuva}\label{kuva6}}
\end{figure}

The approximate analytical theory suggest that to study the relative motion of the satellites
and the correctness of the theory, it is enough to select some small eccentricity and a small
thruster power. Scaling for other values is easy using the theory. In this section we present
some results of numerical experiments and display them  graphically.

We used  for eccentricity the  value $e=.01$ and initial semi-major-axis difference of $1$km.
The size of the radial oscillation is about $2 e a$ as one derives from the
two-body motion and the along-orbit motion  is twice this in the coordinate system in which one
of the satellites is in the origin and the coordinate system rotates with that satellite.


In Fig.(\ref{kuva2}) we compare the oscillations of the $\theta$ angle in a system with all the Earth potential
term [Eq.(\ref{Usarja})] with the model that includes only the $J_2$-term. The time span shown in the figure
corresponds approximately to 1.5 months in the end of a total interval of nearly 2.6 years.
We see that the $\theta$ motion is still  very
similar in both models after this long integration. Thus the effect of the potential terms higher than $J_2$
seems minor at least in the $\theta$ angle.


Figure (\ref{kuva3}) we compare the $\theta$ motion in the full Earth potential and pure two-body motion
for about three moths. There is clear difference in amplitude and also a phase difference. Due to the similarity
of the motions one may conclude that these differences are mainly due to the $J_2$ term that makes the
energy values of the satellites different and the initial derivative values, as computed using the two-body formulae,
somewhat erroneous. In any case the differences are not very large and thus the two-body motion based estimates
remain qualitatively correct.


The Fig.(\ref{kuva4}) compares $\theta$ behavior for a shorter time in the two models as above with the analytical
results following from the Eq.(\ref{thetaeqs}). One can see that the analytical approximation predicts the
period with a few percent accuracy. One concludes that the analytical estimates, although based only on
the Keplerian motion formulae, are useful approximations.


The difference in the relative motion curves in the full perturbation and the two-body
model are compared in Fig.(\ref{kuva5}). Again the differences remain acceptably small.


In the theory the relative eccentricity $\widetilde e=|\fat e_1-\fat e_2|$ is an important quantity.
In Fig.(\ref{kuva6}) we demonstrate that this quantity is almost constant all the time
although the individual eccentricities in the satellite orbits vary considerably.
This result is the same what was observed long ago in quasi-satellite orbits \citep{StabLim}.
This numerical result is quite complicated to prove analytically and we ignore such considerations.


We computed the second derivative of $\langle \Delta^n\rangle$ with respect to $\theta$ for different  powers $n$.
The result is that this quantity is positive for any value of $n$, positive or negative.  Thus one may conclude that
any mutual force between the satellites that are derivable from a power-potential could result into stable
quasi-satellite like motion. In fact there are many other such forces since at least any sum of
power-potentials with positive coefficients would do here. In practice, however, the simple constant force
is probably the most useful for real satellites. The only interesting point here is that in the case of
real world quasi-satellite motion the force is  $r^{-2}$ (plus the indirect term which has only a small effect).
Thus the dynamics with mutual forces, derivable from any power potential, has similar features.
\begin{figure}
\centerline{\includegraphics[width=8.5cm,height=8.0cm]{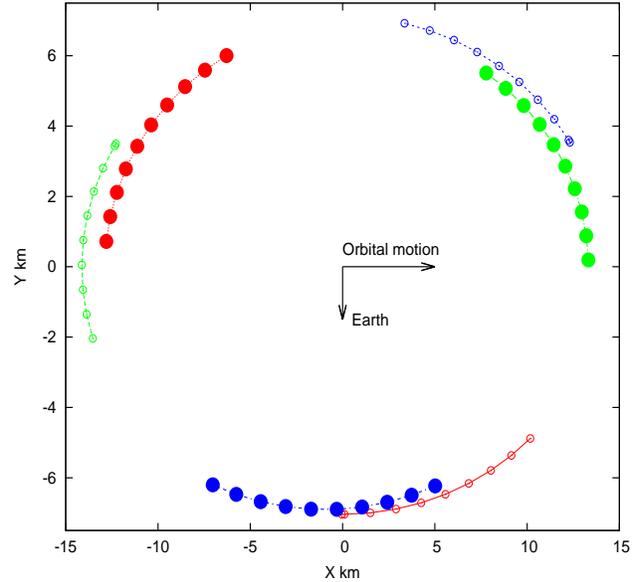}}
\caption{Three satellites keep an almost triangle configuration
from the start to the end of this simulation after about 2.6 years.
In this system the motion perpendicular to the mean orbital plane is only $\le 1\%$
of the size of the system. The colors show the identity of the satellites
and size of dots go from small to bigger according to time.
\label{threesats}\label{kuva12}}
\end{figure}

\begin{figure}
 \includegraphics[width=4.5cm]{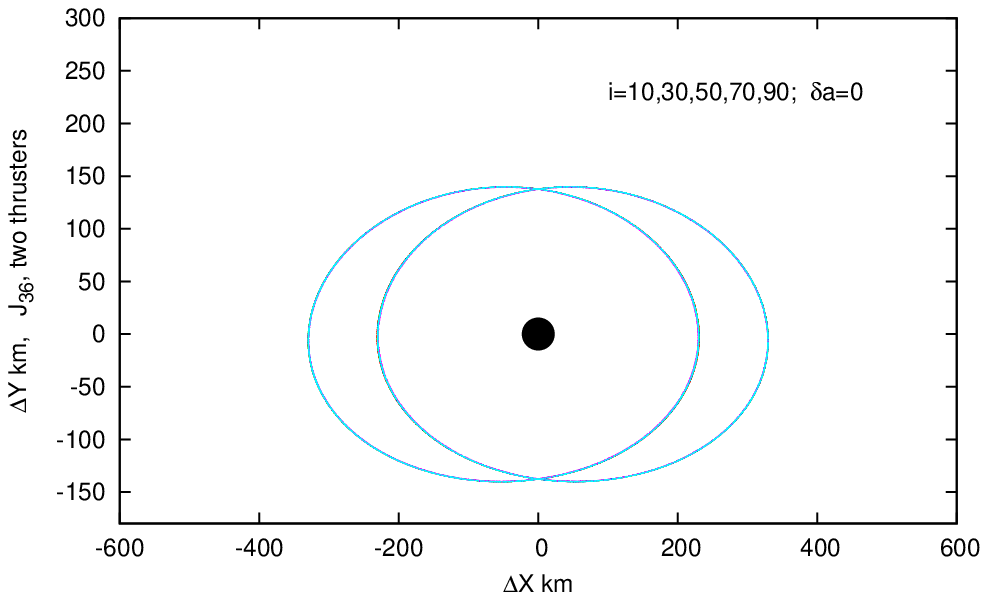} 
 \includegraphics[width=4.5cm]{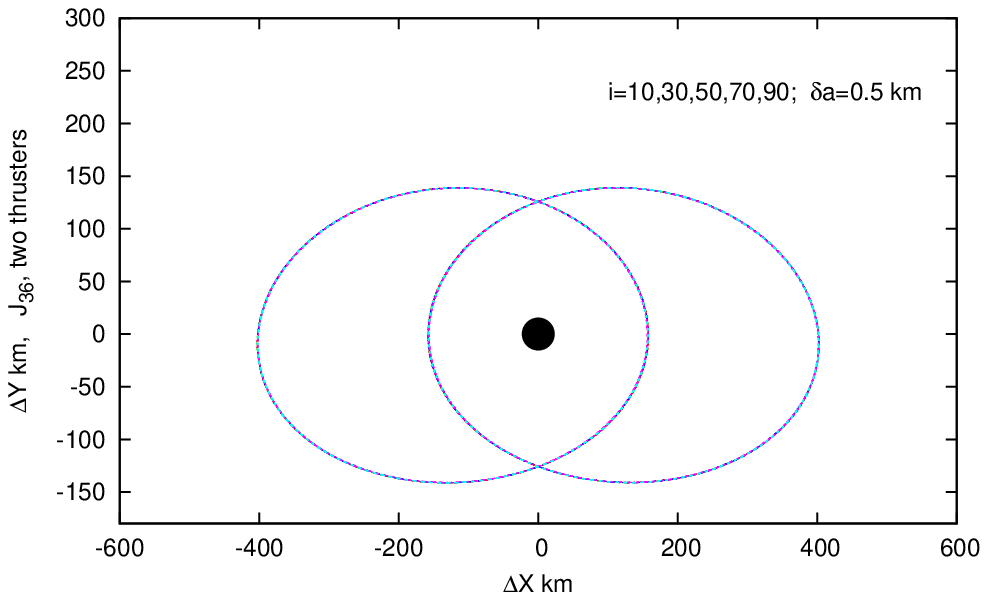} 
 \includegraphics[width=4.5cm]{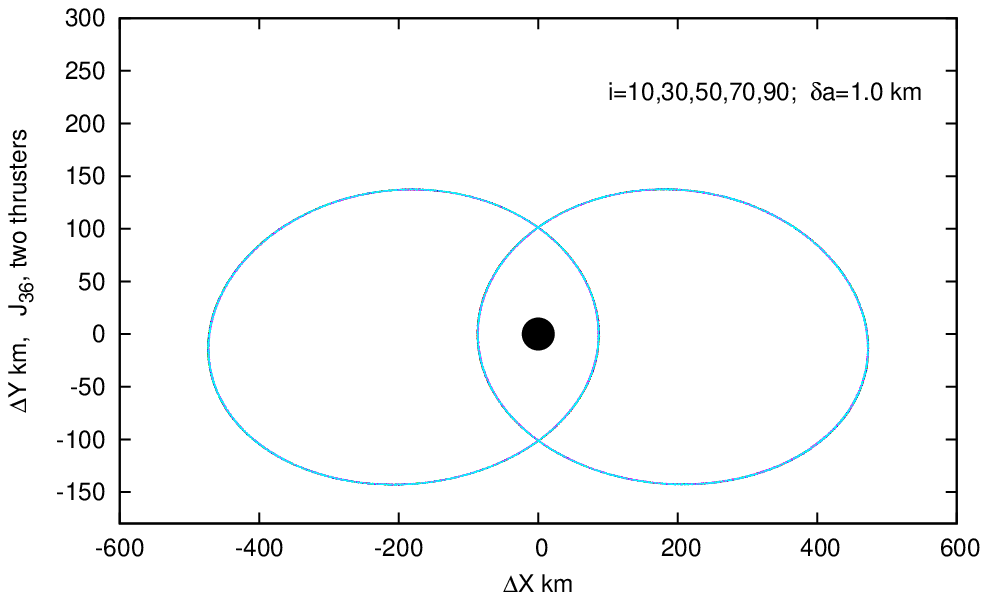} 
\caption{Potential up to $J_{36}$, xy-motion, two thrusters. The number of curves is larger than 1400. \label{RNx}\label{kuva8}}
~

~
\includegraphics[width=4.5cm]{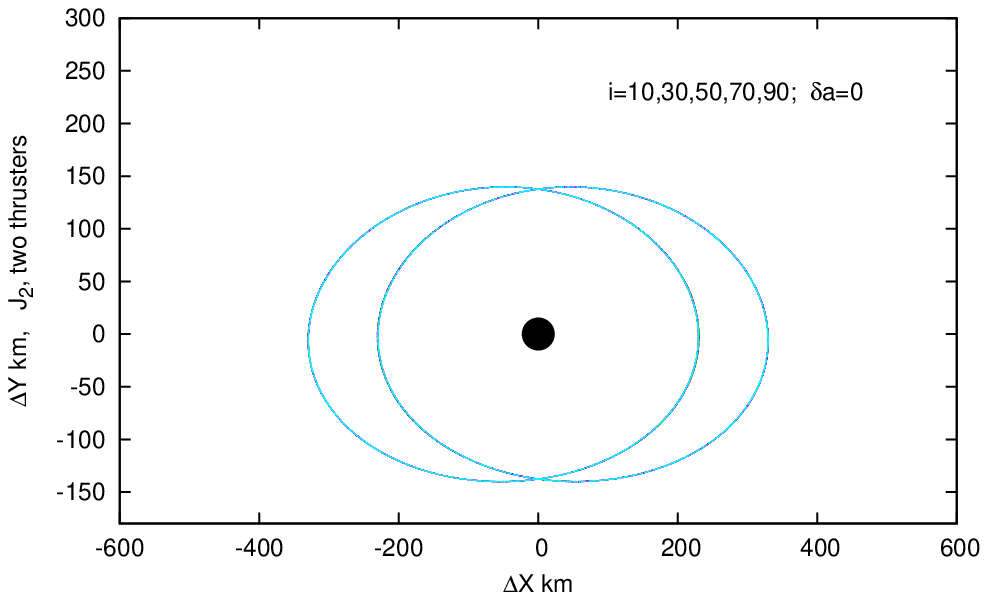} 
\includegraphics[width=4.5cm]{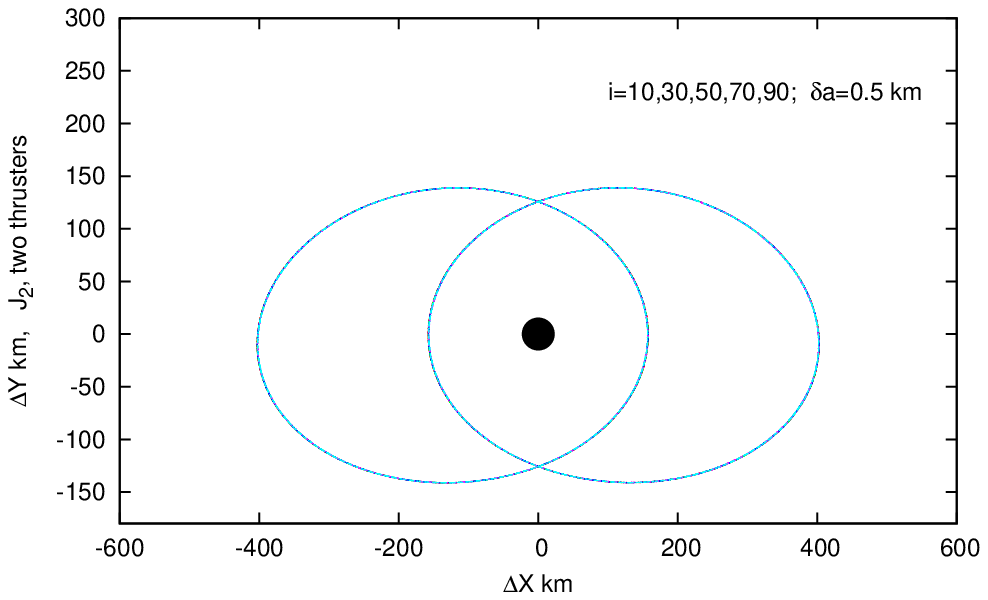} 
\includegraphics[width=4.5cm]{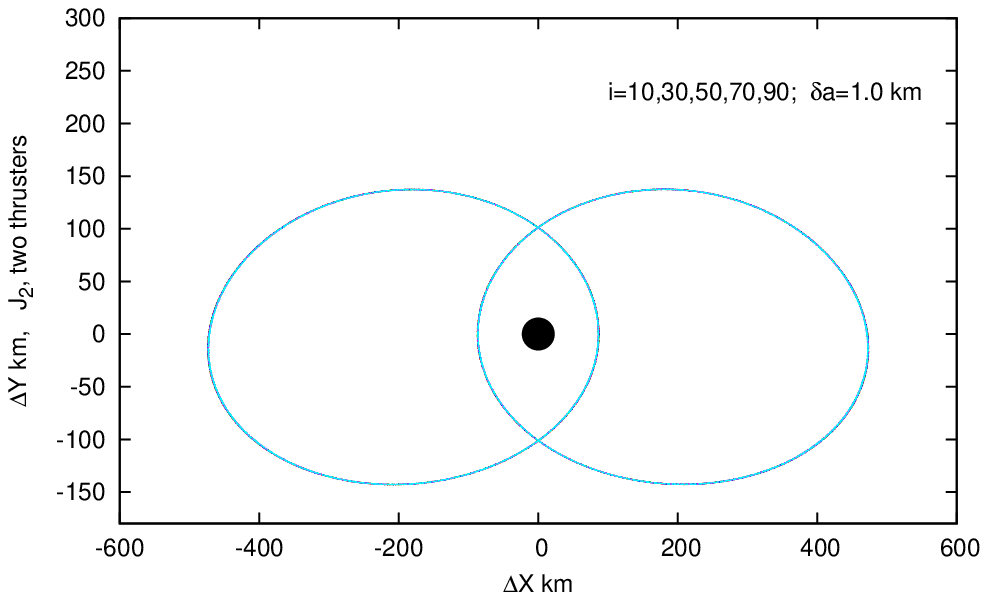} 
\caption{Potential only to $J_2$, xy-motion, two thrusters. The number of curves is larger than 1400.\label{RNJ2}\label{kuva9}}
~

~

 \includegraphics[width=4.5cm]{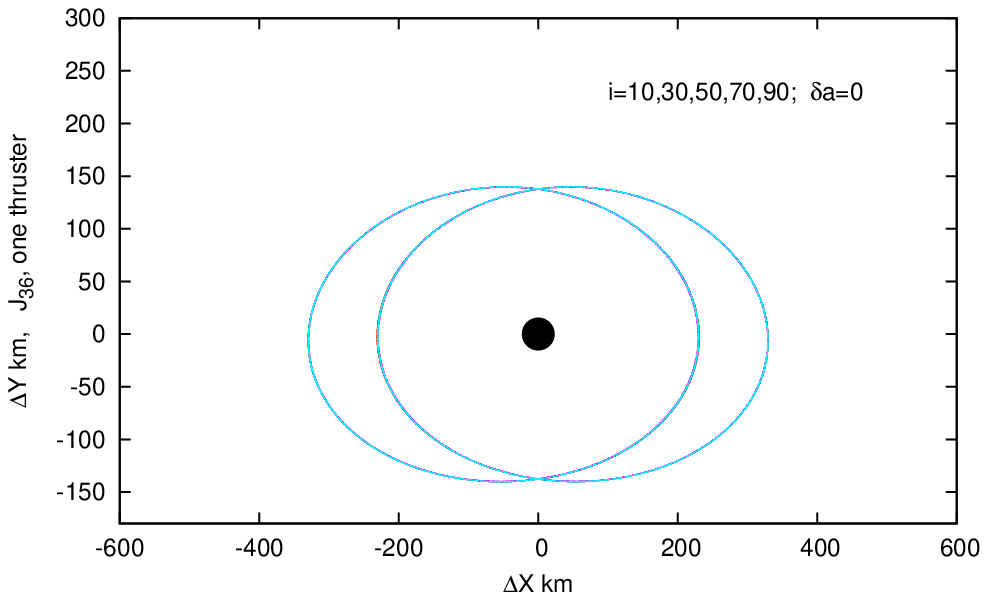}  
 \includegraphics[width=4.5cm]{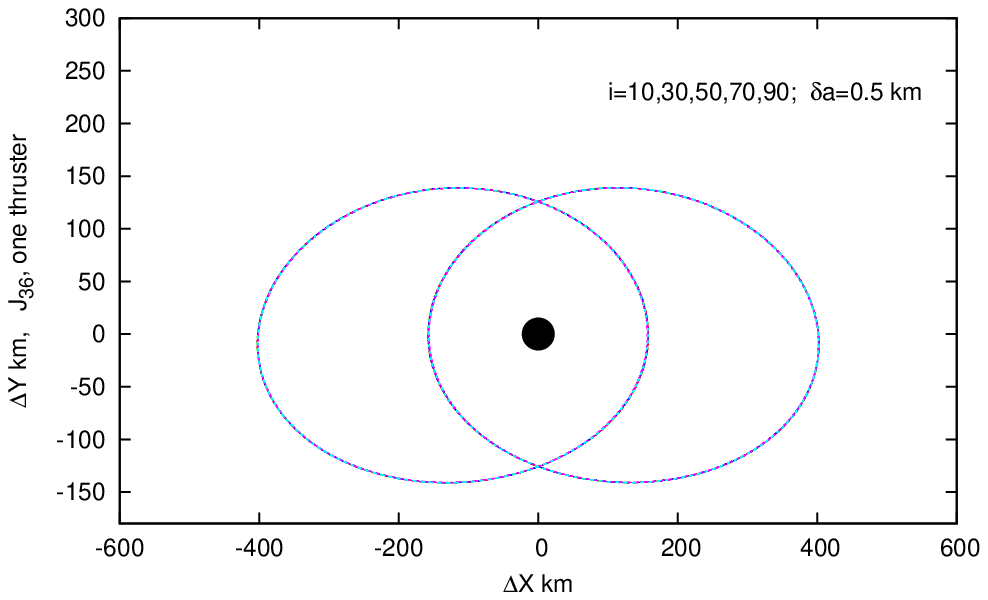}
 \includegraphics[width=4.5cm]{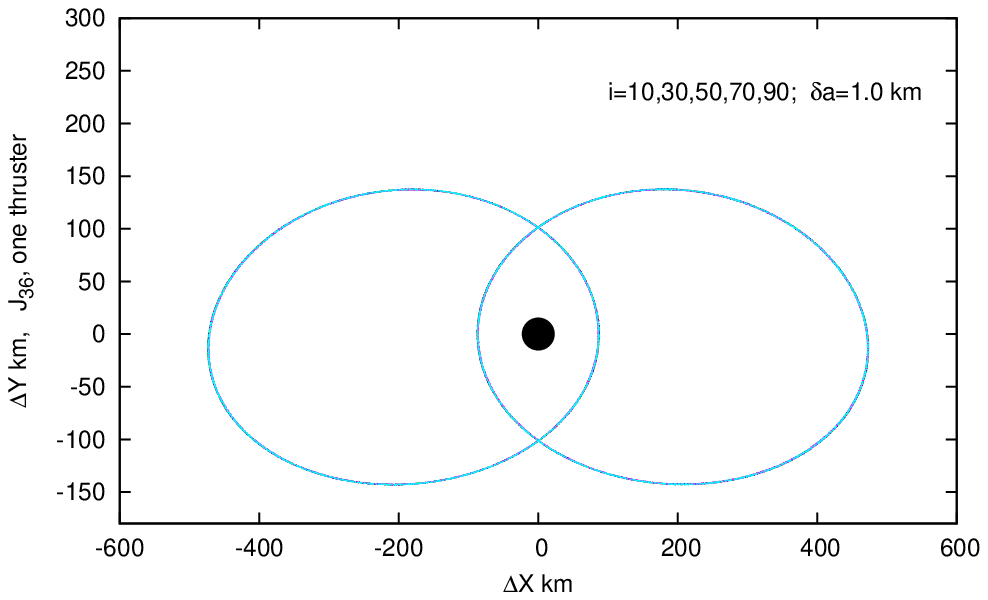}
\caption{Potential up to $J_{36}$, xy-motion, one thruster. The number of curves is larger than 1400.\label{RN1thrus}\label{kuva10}}
~

~

 \includegraphics[width=4.5cm]{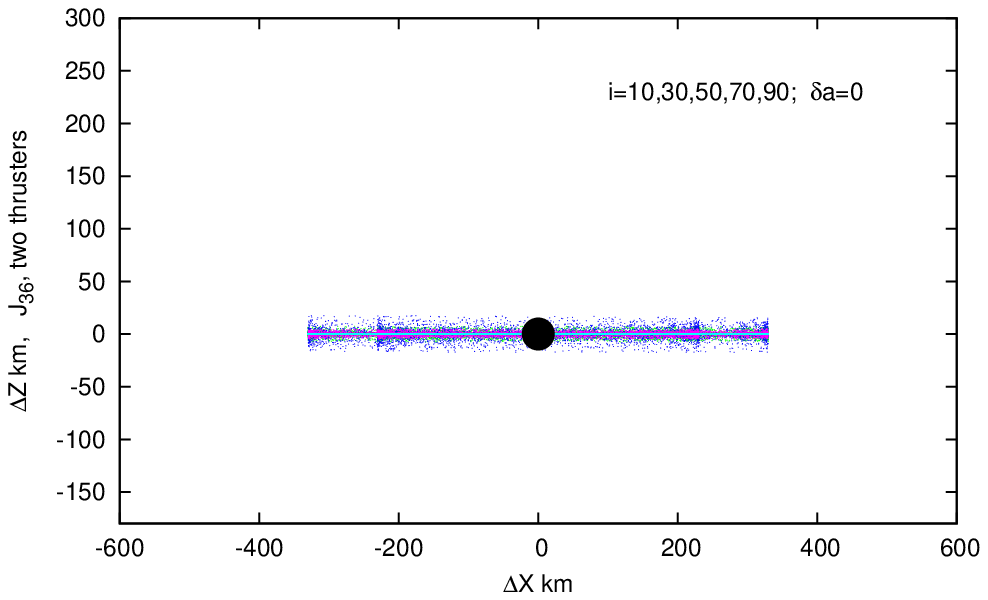} 
 \includegraphics[width=4.5cm]{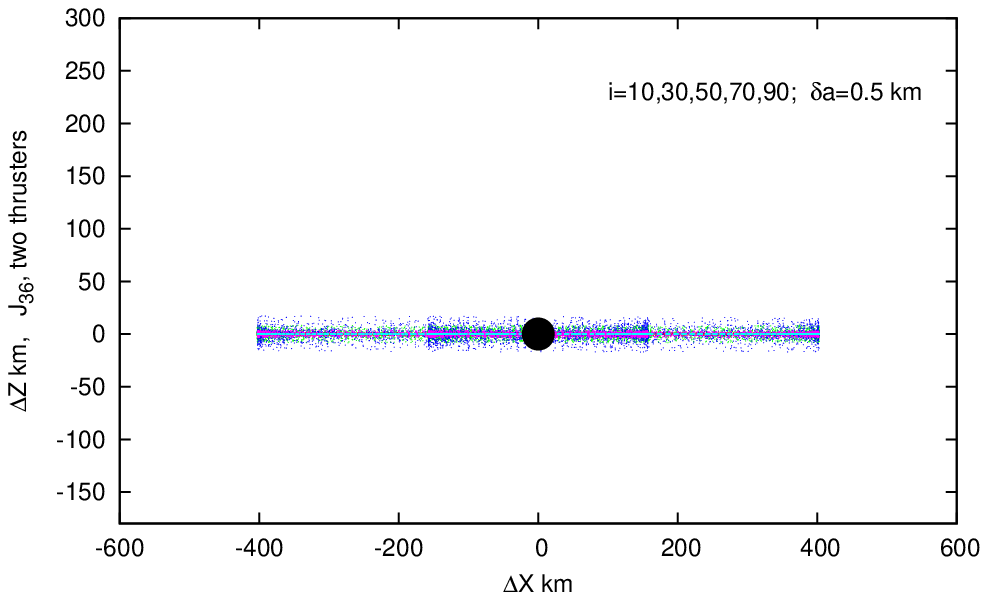}
 \includegraphics[width=4.5cm]{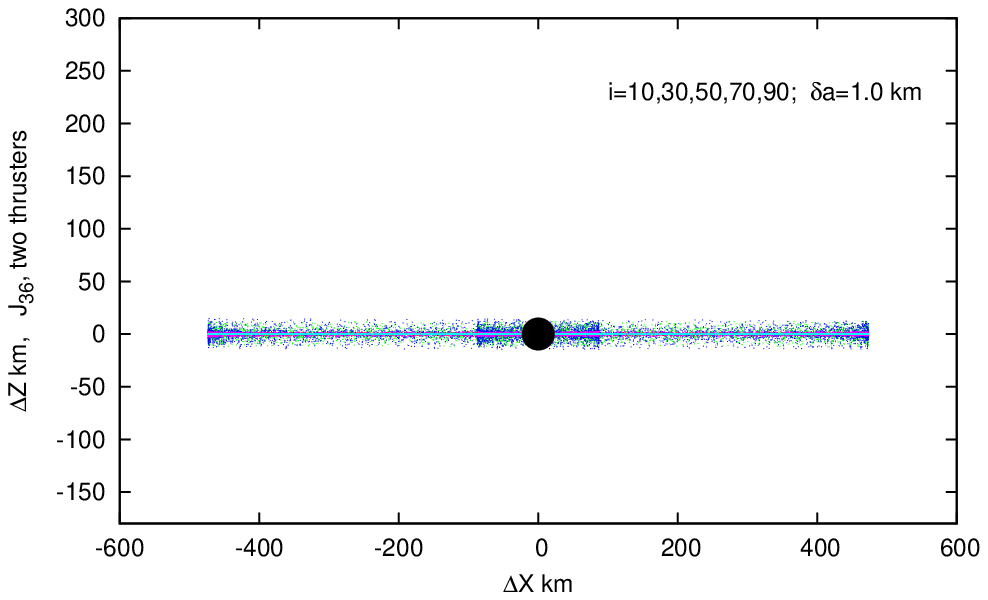}
\caption{Potential up to $J_{36}$, xz-motion, two thrusters. The number of curves is larger than 1400.\label{RNz}\label{kuva11}
 }
\end{figure}

In Figures (\ref{kuva8}) -- (\ref{kuva11}) the relative motions of the satellites are illustrated by plotting it
at every maximum of $|\theta|$ for one period around the Earth. The integration lasted over the time interval
up to $10^5$ time units in our system. This corresponds to about 2.6 years for near Earth satellites.
The results look like just one plotted curve, but are actually more that 1400 curves. Thus one sees the
stability of the system. The results are very similar in the full potential model (marked $J_{36}$,
which includes also Luni-Solar terms) and the $J_2$ models with one or two thrusters. (One thruster
meaning that only one of the satellites has a thruster).
In the Fig.(\ref{kuva11}). the motion with respect to the osculating orbital plane is illustrated
for the full-potential two-thruster computation. Here clearly (like in fact in all the cases) this $xz$-motion
motion is very small compared with the $xy$-motion.

\onecolumn
\twocolumn

\subsection{More satellites}\label{threesat}

For three satellite experiments we used the initial values as explained in the beginning
of this section.
The thruster accelerations for all the satellites were
 directed according to the mean direction of the two other satellites. In some experiments
we randomized these  by modifying the direction cosines randomly by $10\%$ (and then we renormalized
the direction vector). These operations did not have any noticeable effect.

 The result was that the satellites move in a kind of triangle configuration.
Actually the orbits with respect to the center of the triple are approximately ellipses,
but if the scale of the variation of the distance from the Earth is scaled by a factor of $2$
the configuration looks like a circle. This is illustrated in Fig.~\ref{kuva12} where the smaller dots
represent the situation in the beginning and the larger ones show it after about 2.6 years.

Very similar results were  obtained also with more satellites, for example systems of six satellites  seems to behave in a similar way.
Thus it seems possible to keep a cluster of satellites close to each other just by using constant repulsive forces. 
This is probably simpler and other thinkable ways for formation flight of satellites. However this method just provides a way to 
keep satellites in a kind of elliptic like configuration.

\section{Discussion and conclusion}

The motions of the satellites in low Earth orbits are essentially just elliptic motions
so that the relative motions are due to differences in two-body motion.
Our numerical experiments demonstrate that the tandem flight has the effect
that the perturbations from the Earth harmonics, which are orders of magnitude larger than the needed thruster force,
affect the motions of the satellites in the same way and do not make much difference in
their relative motion. This was also confirmed to be the case when high order expansion for
the Earth gravitational potential (equation (\ref{Usarja})) is used.
In other words the relative motion of satellites is mainly due to eccentricities
so that in the orbital coordinate system they seem to move approximately along a small ellipse
in which the axis ratio is $1:2$.
 The only thing the thrusters must do is to keep the mean longitudes nearly same.
For this purpose very small thrusts are enough when the orbits of the satellites
are close to each other. Little differences in the orbits are, however, not important
as we found with simulations.

The method woks as well for two or three or even more satellites.

Surprisingly even  near triangle configuration (in case of three satellites) is
stable in the sense that the relative motion of the satellites keep the same
configuration for very long times.

In our experiments, with the constant acceleration model, the value of relative
acceleration of order $\eps=10^{-6}g$ [g=the standard gravity value ($ \sim 10 {\rm m/sec}^2 $)]  is enough.
Writing this in physical units we can say that
 $\eps\sim 10^{-5} {\rm m}/{\rm sec}^2= .01 {\rm mm}/{\rm sec}^2$  is enough, but
 even   smaller values may do.
Only if the initial difference in semi-major-axis is more than a kilometer
or so, a somewhat stronger force is required, but even in such cases the forces needed
are surprisingly small.
In any case, the required acceleration is easy to find by numerical simulations.
The experiments with the high order expansions, which contains the relevant perturbations
  for the satellite motion,
leads to same results as the ones with only the $J_2$ term, at least within the precision
of our illustrations. This result suggests that the relative motion under the
action of the thruster force is not sensitive to perturbations.
 As mentioned in section \ref{threesat}, even somewhat randomized directions of the
thruster force does not have an effect in the relative motion of the satellites.

One advantage of this method is that the satellites need only information
about the directions of other satellite(s).

{\bf Acknowledgments:} 
Claudiu Prioroc's research has been funded by
the European Commission through the Astrodynamics
Network AstroNet-II, under Marie Curie contract
PITN-GA-2011-289240.

\label{lastpage}

\begin{thebibliography}{99}

\bibitem[Alfriend et al.(2001)]
{AlfEtAl} Alfriend, K.~T.,
Vadali, S.~R., Schaub, H.\ 2001.\ 
Celestial Mechanics and Dynamical Astronomy 81, 57-62.

\bibitem[Beichman et al.(2004)] {BGMR2004} 
 Beichman, C., G{\'o}mez, G., Lo, M., Masdemont, J., Romans, L.\ 2004.\  
 Advances in Space Research 34, 637-644.

\bibitem[Clohessy and Wiltshire(1960)]
{CloWil}Clohessy, W. H. and Wiltshire, R. S., 
Journal of the Aerospace Sciences, Vol. 27, No. 5, 1960, pp. 653–658, 674.

\bibitem[Hill (1878)]{Hill1878}
Hill G.W.,
 American Journal of Mathematics. 1, 1878, 5-26.


\bibitem[Gim and Alfried(2005)]{DiffEquinoctial} 
 Gim, Dong-Woo and Alfriend, K.T.\ 2005 \ 
Celestial Mechanics and Dynamical Astronomy 92, 295–336.
ghEcc

\bibitem[Jackson(1913)]{Jackson} 
 Jackson, J. \ 1913,\ MNRAS, 74, 62-82. 


\bibitem[Kristiansen et al.(2010)]{KPLR}
 Kristiansen, K.~U., Palmer, P.~L., Roberts, M.\ 2010.\ 
 Celestial Mechanics and Dynamical Astronomy 106, 371-390.


\bibitem[Prioroc and Mikkola(2015)]{C-LPSM} 
 Prioroc, C-L., Mikkola, S.\ 2015 \ 
\ New Astronomy 34, 41-46.


\bibitem[Mikkola et al.(2006)]{StabLim}
 Mikkola, S., Innanen, K., Wiegert, P., Connors, M., Brasser, R.\ 2006.\ 
 Monthly Notices of the Royal Astronomical Society
369, 15-24.

\bibitem[Mikkola, Palmer and Hashida (2002)]{MPH02} Mikkola, S., Palmer, P.,
 Hashida, Y.\ 2002.\ 
Celestial Mechanics and Dynamical Astronomy 82, 391-411.

\bibitem[Roscoe et al.(2013)]{HighEcc} 
Roscoe, C.W.T., Vadali, S.R., Alfriend, K.T., Desai, U.P. \ 2013\ 
Acta Astronautica 82, 16–24.


\bibitem[Schaub and Alfriend(2001)]{ShaubAlf2}
 Schaub, H., Alfriend, K.~T.\ 2001.\ 
Celestial Mechanics and Dynamical Astronomy 79, 77-95.

\bibitem[Scheeres,(1998)]{sheeres} Scheeres, D.J. \ 1998.\
Celestial Mechanics and Dynamical Astronomy 70: 75–98.


\bibitem[Vallado and Alfano(2014)]{VA14} Vallado, D.~A., Alfano, S.\ 2014.\ 
Celestial Mechanics and Dynamical Astronomy 118, 253-271.

\end{thebibliography}
\end{document}